\documentstyle [aps,prl,psfig,twocolumn]{revtex}
\begin{document}
\title{ Single-electron soliton avalanches in tunnel junction arrays}

\author{Viktor A. Sverdlov, Daniel M. Kaplan, Alexander N. Korotkov and Konstantin K. Likharev }

\address{
Department of Physics and Astronomy, State University of New York,
Stony Brook, New York 11794-3800 }

\date{\today}
\maketitle
\begin{abstract}
%\setlength{\baselineskip}{16pt}
%\vspace{3cm}
Numerical modeling of correlated single-electron tunneling in uniform 2D arrays of small conducting islands separated by 
tunnel junctions shows the possiblility of soliton-antisoliton avalanches. Though the time duration of any avalanche, and 
the total charge $\Delta Q = ne$ transferred across the array during the avalanche, are always finite, in arrays with 
length $N$ larger than certain critical value $N_c$ and large width $M \gg f(N)$, the avalanche magnitude $n$ may be 
exponentially large, resulting in particular in a giant increase of shot noise. Thermal fluctuations and disorder 
gradually suppress the avalanche effect.  
\end{abstract}

\vspace{1cm}
    

\vspace{1cm}

During the past few years, there was much interest in correlated single-electron 
transport \cite{AveLik,GraDev} in large arrays of small conducting islands separated by tunnel barriers 
- see, e.g., Refs. 
\cite{LikIBM,Bak1Da,DelExp,GeiSch,MooExp,Bak2Da,AvKoNa,DelRev,MooSch,MidWin,Kor94b,Kobaya,Andres,DelHav,Kor96u,Melsen,Cordan,Kurdak,MatLik,KorHop} 
and references therein. This attention is due to several interesting properties 
of the arrays, the most basic of which is the existence of single-electron solitons 
\cite{LikIBM,GeiSch}. Such a soliton consists of an additional single-electron charge placed on an island of the array,
surrounded by a group of totally neutral but strongly polarized islands which screen the electron field at
large distances. Similarly, removal of an electron from a single island creates an ``antisoliton" (i.e., a single-hole
soliton).

The concept of single-electron solitons and antisolitons and their interaction allows a natural explanation of all the 
peculiarities of the array statistics and dynamics, including the Kosteritz-Thouless-like phase transition in 
the 2D case \cite{GeiSch,MooExp,Kobaya}, single-electron oscillations of frequency $f = I/e$ \cite{Bak1Da,DelExp,Kor94b}, 
Coulomb drag \cite{AvKoNa,DelHav}, effects of disorder \cite{MidWin,Kor96u,Melsen,Kurdak}, and shot noise suppression 
\cite{MatLik,KorHop}. The goal of this
Letter is to report the prediction of a new phenomenon in 2D SET arrays: soliton/antisoliton avalanches which lead to 
{\it shot noise enhancement} rather than suppression.

We have obtained theoretical evidence of this effect during Monte Carlo simulation of 2D array
dynamics within the ``orthodox" model of single-electron transport \cite{AveLik}. In this model, which
is quantitatively valid when the tunnel conductances $G$ between the array islands are sufficiently low ($G \ll e^2/\hbar$), single electron 
tunneling events are treated as incoherent transitions with rates
\begin{equation}
\Gamma=\frac{G}{e^2} \frac{\Delta W}{1-\exp(-\Delta W/k_B T)},
\label{eq1}
\end{equation}
where $\Delta W$ is the drop of the 
electrostatic free energy caused by the particular transition. The energy $W$ of a charge configuration was calculated 
within the usual approximation  \cite{LikIBM,GeiSch} (strictly correct for an array sufficiently close to a
conducting ground plane) in which the capacitance matrix includes only diagonal terms $C_0$ and near-diagonal terms $C$ 
which represent island self-capacitances and mutual capacitances between the neighboring islands, respectively. 
We studied not only the average current $I$ through the array, but also the spectral density $S_I(\omega)$ of current 
fluctuations.

The results show that at larger currents the Fano factor $F=S_I(0)/2eI$ decreases to 1/$N$ ($N$ is the longtitudinal number of junctions), just as in 1D arrays
\cite{MatLik}, so that the low-frequency shot noise is indeed suppressed in comparison with the Schottky value $2eI$. 
However, rather unexpectedly, we have found that in uniform arrays at low temperature $T$ and low currents $I$ 
(near the Coulomb blockade threshold) the Fano factor may be much larger than one - see, e.g., Fig.~1. 

\begin{figure}
\centerline{\hbox{
\psfig{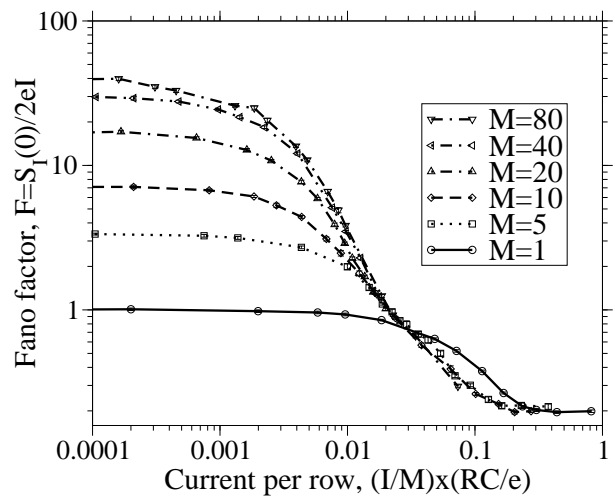}}}
\vspace{0.2cm}
\caption{Fano factor vs.\ dc current for several rectangular arrays of $(N-1)\times M$ similar islands,
with the same length $N=5$ but various widths $M$, for $C_0 = 0$ and $T = 0$. Point size corresponds to 
the Monte Carlo simulation accuracy; lines are only the guides for the eye.}
\label{fig1}
\end{figure}

This fact invites the suggestion that electrons are transferred through the array in ``bundles" with total charge 
$\Delta Q \gg e$; however, since single-electron solitons repel \cite{LikIBM,Bak1Da}, the mechanism
of the bundle formation was not immediately clear. The solution to this puzzle has turned out to be very simple. It
is illustrated by Fig.~2 which shows a few snapshots of a typical Monte-Carlo-simulated process at zero temperature, 
when the dc voltage $V$
applied to the array is just slightly (in this particular case, by $10^{-6} e/C$ ) above the Coulomb blockade threshold $V_t$. At
$V \approx V_t$, the soliton entrance into the initially empty array is the main transport sequence bottleneck, 
and takes
place only after a considerable time pause when the array is free of any extra charges. After the entrance from one electrode, 
the soliton starts to
drift, hopping along the electric field applied to the array (Fig.~2a). When the soliton approaches the opposite electrode, 
its field may 
induce the entrance of one or more antisolitons, i.e. single-hole solitons (Fig.~2b). After these antisolitons have passed 
their transport bottleneck 
at the entrance, they drift mostly along the applied electric field (in the direction opposite to that of the soliton),
the 
attracting field of the initial soliton having little effect on the drift. Because of this, soliton-antisoliton pairs 
frequently miss their chance to recombine, though lateral tunnel junctions between island rows allow such recombination 
and we do observe such events in our simulation. In turn, each
antisoliton approaching the array electrode may trigger the entrance of one or more solitons (Fig.~2c), $\it etc.$ This 
chain reaction results in a soliton-antisoliton avalanche, very much similar to an electric discharge in a gas due
to surface impact ionization \cite{Discha}. Notice that in 1D arrays the soliton and antisoliton cannot pass each other, 
so that the recombination always happens; as a result avalanches are absent and $F\leq 1$ \cite{MatLik}. 

%\vspace{0.4cm}
%\hspace{1.0cm}
%(a)
%\hspace{2.3cm}
%(b)
%\hspace{2.3cm}
%(c)
\begin{figure}[t]
%\vspace{0.1cm}
\centerline{\hbox{
\psfig{figure=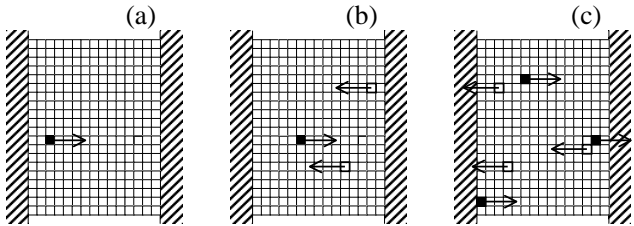,width=8.4cm}}}
\vspace{0.4cm}
%\vspace{0.4cm}
\caption{A typical start of an avalanche in a $15\times 20$-island array biased by dc voltage $V\simeq V_t\simeq 2.35e/C$. (The voltage is 
applied between the array electrodes shown as hatched rectangles.) Black squares mark the islands with an extra electron (the single-electron soliton centers), and white squares those with an extra hole (the antisoliton centers). Arrows show the directions of soliton and antisoliton
motion. $C_0/C = 0.01$, $T = 0$. (a) $t = 8.48$, (b) $t = 30.6$, (c) $t = 119$; here $t$ is time measured 
(in the units of $C/G$) from the (random) moment of the first soliton entrance.}
\label{fig2}
\end{figure}

Due to the stochastic character of transport within the framework of the orthodox theory, there is always a chance that every soliton and 
antisoliton leaves the array {\it without} triggering the entrance of any solitons of the opposite charge. Because of this,
the avalanches always have a finite duration in time and finite ``magnitude" $n$ (the latter may be defined by the 
equation 
$\Delta Q = ne$ for the total charge transferred through the array during one avalanche). Nevertheless, the avalanches 
may be rather large: in wide arrays we have observed $n$'s up to $4\times 10^{6}$, limited only by the computer simulation time.

In the range where avalanches are distinct, computation of $F$ may be sped up 
considerably by using formulas for ``charge blocks", which were derived in Ref. \cite{Kor94a} (in a different context):

\begin{eqnarray}
&& S_I(0)=\frac{2}{\langle\tau\rangle}\left( e^2\langle n ^2\rangle + I^2\langle\tau ^2\rangle -2eI\langle n\tau \rangle\right) , 
\label{Idep1}\\
&& I=e\langle n\rangle /\langle \tau\rangle .
\label{Idep2}
\end{eqnarray}

If avalanches do not overlap, $n$ is just the magnitude of a single avalanche, while
$\tau$ is the interval between the beginnings of the adjacent avalanches. In the particular case $I \rightarrow 0$, Eqs. (\ref{Idep1}) and (\ref{Idep2}) yield

\begin{equation}
F=\frac{\langle n^2\rangle }{\langle n\rangle }.
\label{F0}
\end{equation}

Figures 1 and 3 show the dependence of $F$ on the main parameters of the system: array length $N$, width $M$, island
capacitance ratio $C_0/C$, and applied voltage $V$ [the last dependence is presented parametrically via the induced dc 
current $I(V)$].  Most of these dependences may be readily understood, at least qualitatively, using the avalanche picture
discussed above. For example, if the ratio $C_0/C$ is increased, the soliton radius decreases \cite{LikIBM,GeiSch}, so 
that the soliton-antisoliton interaction is quenched, and $F$ is decreased. Larger $N$ gives the soliton 
more time to induce solitons of the opposite sign, so that $F$ grows. Finally, in very narrow arrays (small $M$) the 
soliton-antisoliton recombination suppresses the avalanche magnitude, so that the Fano factor also decreases.

\begin{figure}
\centerline{\hbox{
\psfig{figure=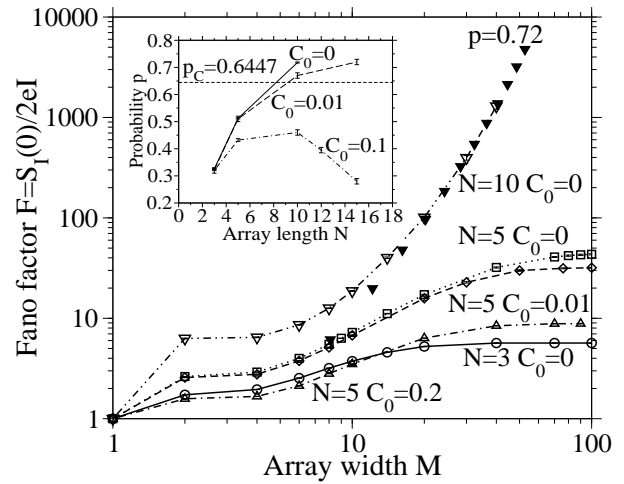,width=8cm}}}
\vspace{0.3cm}
\caption{Fano factor at $I \rightarrow 0$ ($V \rightarrow V_t$) as a function of the array width, for $T = 0$ and 
several values of $N$ and $C_0/C$.
Open points: Monte Carlo simulation of the initial problem. Closed triangles: Monte Carlo simulation of the ``macromodel".  Inset: Parameter $p$ of the ``macromodel'' as a function of $N$ and $C_0/C$. }
\label{fig3}
\end{figure}

We have found that the avalanche statistics dependence on $M$ in the opposite limit of large $M$ may be understood as
follows. In wide arrays, the probability $p$
that a soliton triggers the entrance of an antisoliton, and does not recombine with it, should not depend much on the array 
width $M$. The average number $f$ of rows separating the soliton and the neighboring
antisoliton, may depend on $N$ and $C_0/C$, but should be also virtually independent of $M$. This is why we may 
introduce an approximate ``macromodel" which is schematically shown in the inset of Fig.~4: an array of width $M$ is 
presented as a parallel connection 
of $m = M/f$ channels of equal width $f = f(N, C_0/C)$.
We break the avalanche into time stages of equal duration and assume that a passage of a soliton in some 
channel at stage $i$ triggers the antisoliton passage in each of the
two neighboring channels at stage $(i+1)$ with probability $p = p(N, C_0/C)$. The lateral (open) sides of the array are described by setting the 
corresponding probability to 0. Mathematically, our macromodel is exactly the problem of the directed 
bond percolation problem on a 2D square lattice \cite{Percol} within a stripe of width $m$.

\begin{figure}
\centerline{\hbox{
\psfig{figure=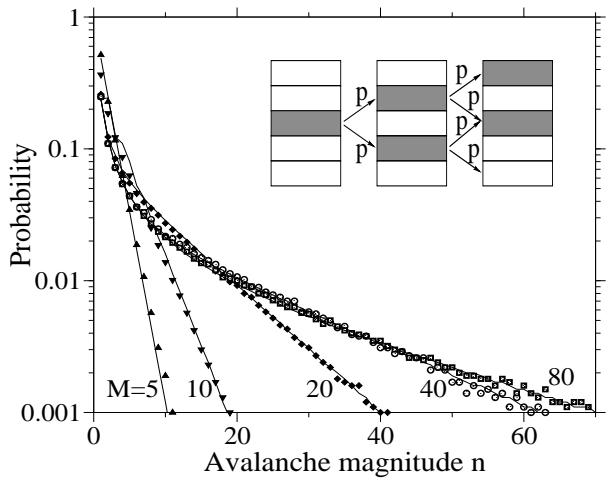,width=8cm}}}
\vspace{0.3cm}
\caption{Avalanche magnitude hystogram for several arrays of different widths $M$, all with $N = 5$ and $C_0 =0$. Points: Monte Carlo simulation of the initial problem. Solid lines: Monte Carlo simulation of the ``macromodel" 
(shown in the inset) for $p$ = 0.512 and $f$ = 3.}
\label{fig4}
\end{figure}  

Despite the approximate character of the macromodel, it allows the results of the initial problem to be reproduced
remarkably well. For example, Fig.~4 shows avalanche amplitude histograms (probability to find an avalanche with particular magnitude $n$) for 5 arrays
of different widths. The Monte Carlo simulation results virtually coincide with the macromodel results, despite
only two parameters ($p$ and $f$) were available for fitting all 5 curves. (An equally good fit was obtained for all 
other values of $N$ and $C_0/C$ we have studied if $M$ was large enough, $m \gg 1$.) At the same time, the macromodel is much faster for simulation, so that
with the same computer resources, results may be obtained in a much broader range of array parameters. For example, 
open points in Fig.~3 show Monte Carlo results for the initial model; for $N = 10$, we could hardly get 
acceptable accuracy at $M>40$. For the macromodel described above, calculation of $F$ in that point 
(with the same error bars) took 35 times less CPU time, and we could continue calculations all the way up to $M$ = 55
(closed triangles).

The macromodel also gives a clear explanation why the growth of the avalanche magnitude with array width $M$ saturates 
for shorter arrays, but is unbound for longer arrays (Fig.~3). The directed bond
percolation problem on square lattice has a percolation threshold $p_c=0.6447$, beyond which
there is a finite probability of having an infinite percolation cluster on an infinite lattice. We have found 
that for our problem, $p$ becomes larger than $p_c$ if the array length is above a certain value
$N_c$. ($N_c \approx 8$ for $C_0 = 0$, and grows with the increase of the $C_0/C$ ratio; for large enough $C_0/C$, $p$ never reaches $p_c$ at all -- see inset in Fig.~3.) This means that at $N>N_c$ and 
unlimited $m$ there is a finite probability of having an infinite avalanche which, once started, would never end. The large, but finite, array width (and hence a large but finite $m \gg 1$) stops the avalanche growth in time, and limits its 
magnitude at a finite (though exponentially large in $m$) level. 

The avalanche effect is most strongly expressed in uniform arrays at zero temperature, while both thermal fluctuations and
array disorder gradually suppress it. Figure~5 shows a typical dependence of the Fano factor on the applied voltage
for several values of temperature.  For $T = 0$, transport is possible only above the Coulomb blockade threshold
$V>V_t$, so that below this point $F$ is undetermined. However, even very small temperature fluctuations reveal the
second branch of this dependence, since the initial energy barrier for soliton entrance to the array may be overcome by
thermal activation, so that dc current becomes finite (though may be very small). Figure 5 shows that for $V<V_t$, 
the Fano factor decreases as applied voltage is decreased,  approaching 1 for smaller $V$.  The 
reason for this decrease is that as $V \rightarrow 0$, it is harder and harder for the initial soliton to trigger the
antisoliton entrance, so that single-soliton passages become the dominant component of transport. (At very small voltages, 
$eV/2<k_BT$, the Fano factor starts to grow again as $2k_BT/eV$ due to quasi-equilibrium thermal fluctuations. However,
in our case of low temperatures, $k_BT \ll eV_t$, this growth corresponds to exponentially low dc current.) Larger 
temperature leads to gradual de-correlation of the moments when solitons and antisolitons enter the array, and hence to 
a gradual suppression of the avalanches which depend on this correlation.

\begin{figure}
\centerline{\hbox{
\psfig{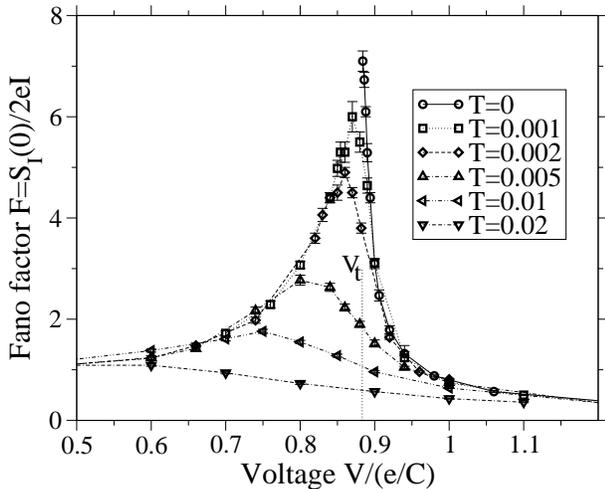}}}
\caption{Fano factor as a function of applied voltage $V$ for an array with $N = 5$, $M = 10$ and $C_0 = 0$, for several 
temperatures (measured in $e^2/k_B C$). $V_t = 0.883e/C$ is the Coulomb blockade threshold.}
\label{fig5}
\end{figure}

The effect of disorder is qualitatively similar (in qualitative agreement with the results for the directed bond
percolation \cite{Noest1}). For example, in a $5\times 20$ junction array with $C_0 = 0$, the introduction of background charges 
$Q_0 = 0.5e$ to just 6 percent of the islands at random usually suppresses avalanches completely, reducing $F$ to 1.  

To summarize, we have found strong theoretical evidence for (and a simple explanation of) a new effect of 
single-electron-soliton and antisoliton avalanches in 2D arrays of small conducting islands. This effect arises due to 
soliton-assisted entrance of antisolitons into neighboring, but different rows of the array, thus avoiding recombination. 
The basic physics of the effect is well captured by a simple ``macromodel", equivalent to the directed percolation model 
on a square-lattice strip of finite width. Thermal fluctuations and array disorder suppress the avalanches. Presently, we 
are working toward a quantitative characterization of the latter suppression, in particular with the goal to evaluate 
feasibility of experimental observation of the soliton avalanches using various techniques of array fabrication 
\cite{DelExp,MooExp,Kobaya,Andres,Cordan,Kurdak}. Apparently, this may be easier to accomplish in an array with 
tunnel-uncoupled rows (essentially, a generalization of the 2-row system studied earlier \cite{AvKoNa,DelHav}), where 
tuning of dc voltages applied to each row may be sufficient to compensate for a moderate disorder.

The work was supported in part by the Engineering Research Program of the Office of Basic Energy Sciences at the 
Department of Energy.


\begin{thebibliography}{99}

\bibitem{AveLik} D.V.~Averin and K.K.~Likharev, in {\it Mesoscopic Phenomena 
in Solids}, ed. by B.~Altshuler {\it et al.} (Elsevier, Amsterdam, 1991), p. 173.
\bibitem{GraDev} {\it Single Charge Tunneling}, ed. by H.~Grabert and M.H.~Devoret (Plenum, New York, 1992).
\bibitem{LikIBM} K.K.~Likharev, IBM J. Res. Devel. {\bf 32}, 144 (1988).
\bibitem{Bak1Da} N.S.~Bakhvalov, G.S.~Kazacha, K.K.~Likharev, and S.I.~Serdyukova, Sov. Phys. JETP {\bf 68}, 581 (1989).
\bibitem{DelExp} P. Delsing, K.~Likharev, L.~Kuzmin, and T.~Claeson, Phys. Rev. Lett. {\bf 63}, 1861 (1989).
\bibitem{GeiSch} U. Geigenm\"uller and G. Sch\"on, Europhys. Lett. {\bf 10}, 765 (1989).
\bibitem{MooExp} J.E.~Mooij, B.J.~Van~Wees, L.J.~Geerligs, M.~Peters, R.~Fazio, and G. Sch\"on, Phys. Rev. Lett. {\bf 65}, 645 (1990).
\bibitem{Bak2Da} N.S.~Bakhvalov, G.S.~Kazacha, K.K.~Likharev, and S.I.~Serdyukova, Physica B {\bf 173} 581 (1991).
\bibitem{AvKoNa} D.V.~Averin, A.N.~Korotkov, and Yu.V.~Nazarov, Phys. Rev. Lett. {\bf 66}, 2818 (1991).
\bibitem{DelRev} P. Delsing, in {\it Single Charge Tunneling}, (Ref. 2), p.249.
\bibitem{MooSch} J.E.~Mooij, G.~Sch\"on, in {\it Single Charge Tunneling}, (Ref. 2), p. 275.
\bibitem{MidWin} A.A.~Middleton and N.S.~Wingreen, Phys. Rev. Lett. {\bf 71}, 3198 (1993).
\bibitem{Kor94b} A.N.~Korotkov, Phys.~Rev. B {\bf 50}, 17674 (1994).
\bibitem{Kobaya} S.~Kobayashi, A. Kanda, and R. Yamada, Jpn. J. Appl. Phys. {\bf 34}, 4548 (1995).
\bibitem{Andres} R.P.~Andres {\it et al.}, Science {\bf 273} 1690 (1996).
\bibitem{DelHav} P.~Delsing, D.B.~Haviland, and P.~Davidsson, Czech. J. Phys. {\bf 46}, 2359 (1996).
\bibitem{Kor96u} A.N.~Korotkov, unpublished (1996); see Fig.~13 in K. Likharev, Proc. IEEE {\bf 87}, 606 (1999).
\bibitem{Melsen} J.A.~Melsen, U.~Hanke, H.-O.~M\"uller, and K.-A. Chao, Phys. Rev. B {\bf 55}, 10642 (1997).
\bibitem{Cordan} A.S.~Cordan, A.~Goltzene, and H.~Launois, J. Appl. Phys. {\bf 84} 3756	(1998).
\bibitem{Kurdak} C.~Kurdak, A.J.~Rimberg, T.R.~Ho, and J.~Clarke, Phys. Rev. B {\bf 57}, R6842 (1998).
\bibitem{MatLik} K.A.~Matsuoka and K.K.~Likharev, Phys. Rev. B {\bf 57}, 15613 (1998).
\bibitem{KorHop} See Appendix in A.N.~Korotkov and K.K.~Likharev, Phys. Rev. B {\bf 61}, 15975 (2000).
\bibitem{Discha} See, e.g., R.W.~Crompton, {\it Gaseous Electronics and its Applications} (Kluwer, Dordrecht, 1991). 
\bibitem{Kor94a} A.N.~Korotkov, Phys.~Rev. B {\bf 49}, 10381 (1994).
\bibitem{Percol} See, e.g., {\it Percolation Structures and Processes}, ed. by G. Deutsher {\it et al.} 
(Hilger, Bristol, 1983).
\bibitem{Noest1} A.J.~Noest, Phys.~Rev.~Lett. {\bf 57}, 90 (1986).
\end{thebibliography}
\end{document}